\documentclass[12pt,english,floatfix,nofootinbib,superscriptaddress,aps,prd,preprint]{revtex4}
\usepackage[utf8]{inputenc}
\usepackage{float}
\usepackage{array}
\usepackage{lipsum}
\usepackage{graphicx}
\usepackage{amsmath,amsthm,amsfonts,amssymb}
\usepackage{graphicx}
\usepackage[english]{babel}
\usepackage{color}
\usepackage{tensor}
\usepackage{esint}
\usepackage[dvips]{epsfig}
\usepackage[dvips]{graphicx}
\usepackage{float}
\usepackage{units}
\usepackage{textcomp}
\usepackage{mathrsfs}
\usepackage{amsmath}
\usepackage[makeroom]{cancel}
\usepackage{amssymb}
\usepackage{amsbsy}
\usepackage{amsfonts}
\usepackage{amssymb,mathrsfs,xcolor}
\usepackage{esint}
\usepackage{braket}
\usepackage{array}
\usepackage{graphicx}

\makeatletter

\makeatletter\usepackage{babel}

\usepackage{hyperref}
\hypersetup{
    colorlinks,
    citecolor=blue,
    filecolor=green,
    linkcolor=purple,
    urlcolor=red,
}

\usepackage{slashed}

\newcommand{\ie}{\begin{equation}}
\newcommand{\fe}{\end{equation}}
\newcommand{\se}{\begin{eqnarray}}
\newcommand{\ff}{\end{eqnarray}}

\begin{document}

\title{Lorentz-violating scenarios in a thermal reservoir}


\author{A. A. Ara\'{u}jo Filho}
\email{dilto@fisica.ufc.br}
\affiliation{Universidade Federal do Cear\'a (UFC), Departamento de F\'isica,\\ Campus do Pici, 
Fortaleza - CE, C.P. 6030, 60455-760 - Brazil.}


\date{\today}

\begin{abstract}

In this work, we analyze the thermodynamic properties of the graviton and the generalized model involving anisotropic Podolsky and Lee-Wick terms with Lorentz violation. We build up the so-called partition function from the accessible states of the system seeking the following thermodynamic functions: spectral radiance, mean energy, Helmholtz free energy, entropy and heat capacity. Besides, we verify that when the temperature rises, the spectral radiance $\chi(\nu)$ tends to attenuate for fixed values of $\xi$. Notably, when parameter $\eta_{2}$ increases, the spectral radiance $\bar{\chi}(\eta_{2},\nu)$  weakens until reaching a flat characteristic. Finally, for both theories, we perform the calculation of the modified black body radiation and the correction to the \textit{Stefan–Boltzmann} law in the inflationary era of the universe.

\end{abstract}

\maketitle

\section{Introduction}

In theoretical physics, there exists a memorable problem which is putting on an equal footing the so-called Standard Model \cite{schwartz2014}, provided by consistent experimental data in predicting the behavior of fundamental particle physics, and the widespread General Relativity \cite{wald2010}, which has the purpose of regarding gravity as a geometric theory. Since all these approaches are intensively well tested, if there exists a conciliation for both, one will expect a unique and fundamental theory of quantum gravity \cite{rovelli2004}. Moreover, this structure could bring about the feasibility of investigating some new phenomena not yet individually described by them. Nevertheless, up to now, there are neither experimental nor observational indications of any fingerprints of such a unified theory perhaps due to the fact that its effects are highlighted when the energy range around the Planck mass, i.e., $m_{\mathrm{P}} \sim 10^{19}$ GeV, is taken into account.

Nowadays, since it is impossible to have the access of such scale, a reasonable way of working on it has been developed considering the viewpoint that quantum gravity phenomena can be recognized by the proliferation of their effects at attainable energies. In this sense, one of the most remarkable possibilities regards the violation of Lorentz symmetry. Supporting such theory, there are many different mechanisms that bring out Lorentz-violating effects such as in string theory \cite{mavromatos2007}, Horava-Lifshitz gravity \cite{pospelov2012}, noncommutative field theories \cite{carroll2001} and loop quantum gravity \cite{alfaro2002}.

Having been proposed about thirty years ago by Kostelecký \textit{et al.}, the Standard Model Extension (SME) \cite{k1,k2,k4,k5,k7} is an extended version for the usual Standard Model theory. It possesses Lorentz-violating terms, which are rather tensor terms acquiring a nonzero vacuum expectation value, coupled with physical fields preserving their coordinate invariance and a violation of Lorentz symmetry when particle frames are considered \cite{kostelecky2001}. Likewise, such theoretical background has been the precursor for an expressive number of works involving the fermionic sector \cite{f1,f2,f3,f4,f5}, the electromagnetic CPT-odd and Lorentz-odd term \cite{e1,e2,e3,e4,e5,e6} as well as the CPT-even and Lorentz odd gauge sector \cite{c1,c2,c3,c4}.

Over the last years, the connection between Lorentz violation and theories including higher derivative operators has received much attention \cite{d1,d2,d3,schreck2014,cuzinatto2011,casana2018,adailton,anacleto2018,borges2019}. As a matter of fact, it may have operators of higher mass dimensions incorporating for instance higher-derivative terms. Being in contrast to the minimal version of the Lorentz violating extensions, its nonminimal approach has the advantage of possessing an indefinite number of such contributions \cite{schreck2014}. In this sense, the latter version of the SME was first proposed considering both the photon \cite{kostelecky2009electro} and the fermionic sectors \cite{kostelecky2013fermions}.

In this direction, the first illustration of a higher-derivative electrodynamics was proposed by
Podolsky \cite{podolsky1942} having a noticeable feature which is the generation of a massive mode without losing the gauge symmetry. In that paper, it was initially studied the gauge-invariant dimension-6 term, $\theta^{2}\partial_{\alpha}F^{\alpha\beta}\partial_{\lambda}F\indices{^\lambda_\beta}$, with a coupling constant $\theta$, which afterward would be known as the Podolsky parameter, with the mass dimension being $-1$. Clearly, such theory displays two distinct dispersion relations, i.e., the usual massless mode and the massive mode which possesses the advantage of avoiding divergences ascribed to the pointlike self-energy. Nevertheless, considering the quantum level, the latter mode gives rise to the appearance of ghosts \cite{accioly2010}.

Additionally, in the late 1960s, there exists another noteworthy extension of Maxwell theory with higher derivatives being described by the dimension-6 term $F_{\mu\nu}\partial_{\alpha}\partial^{\alpha
}F^{\mu\nu}$, the Lee-Wick electrodynamics \cite{lee1969,lee1970}. Notably, this theory leads to a finite self-energy for a pointlike charge in $(1 + 3)$ spacetime dimensions and to the appearance of a bilinear contribution to the Maxwell Lagrangian. This is analogous to the Podolsky term showing an opposite sign though.
Such “incorrect” sign outputs energy instabilities at the classical level, whereas it brings out a negative norm states in the Hilbert space at the quantum level. Moreover, it was also Lee and Wick who first proposed a mechanism seeking the preservation of unitarity by removing all states with negative norm from the Hilbert space. In the last decade, this theory came back to obtain notability with the proposal of the Lee-Wick Standard Model \cite{chivukula2010,underwood2009,grinstein2008}, based on non-Abelian gauge structure free of quadratic divergences. Such model was widespread having many contributions for both theoretical and phenomenological approaches \cite{carone2009,alvarez2009}. Furthermore, it is worth mentioning that, in this general context, investigations of Lorentz violation extensions are discussed \cite{turcati2014} including applications to the interaction of pointlike and spatially extended sources \cite{accioly2014stra,barone2015}.

On the other hand, the focus on an extension of the SME, which takes into account gravity, arouses from the fact that the Lorentz violation can be expected to be a key element for a quantum theory of gravitation. As a matter of fact, it is important to note that Lorentz-violating effects might be expressive in regions where the curvature or torsion are accentuated, as in the vicinity of black holes for instance \cite{barausse2013}. Besides, these implications can represent a notable role in cosmological scenarios being illustrated by either dark energy \cite{peebles2003} or dark matter \cite{arkani2009}. Moreover, there are others whose anisotropy factors can be added in the Friedman-Robertson-Walker solutions \cite{hawking1973}. In addition,
the main motivation of constructing a theory consistent with gravity, i.e., being in agreement with Bianchi identities and so forth, is having a consistent formalism seeking to maintain the local observer Lorentz covariance, despite the presence of local particle Lorentz violation \cite{kostelecky2004}. In this way, it is worth pointing out that there are investigations regarding Lorentz violation in the linearized gravity \cite{ferrari2007} and others \cite{boldo2010,pereira2011}.

Furthermore, up to now, there are many works based on the analysis of the thermodynamic functions in the gravitational scenario mainly regarding the Cosmic Microwave Background (CMB) \cite{magueijo1994cosmic,cillis1996photon,chen1995resonant,ejlli2013,chen2013}. However, there are very few studies of such thermodynamic properties ascribed to the linearized theory of gravity within the context of Lorentz violation. In this sense, our starting point is taking into account a similar analysis encountered in Refs. \cite{magueijo1994cosmic,cillis1996photon,chen1995resonant,ejlli2013,chen2013}, but proposing rather the inflationary epoch of the universe, i.e., $\beta=1/\kappa_{B}T=10^{-13}$ GeV, within the context of Lorentz violation to utilize the modification from the black body radiation spectra as well as from the \textit{Stefan-Boltzmann} law as an alternative to investigate cosmological scenarios. Besides, for the sake of giving a complement to this analysis, we provide the calculation of Helmholtz free energy, mean energy, entropy and heat capacity.

In addition, there is a lack in the literature ascribed to the investigation of the thermodynamic properties for the generalized anisotropic Podolsky with Lee-Wick terms. In this viewpoint, it is noteworthy to accomplish such analysis in order to verify how the modified massless mode behaves to perhaps reveal new phenomena which might be applied to either condensed matter or statistical thermal physics.


\section{Graviton with Lorentz violation}

In this section, we begin with the action responsible for the dynamics of the bumblebee field $B_{\mu}$ written as
\ie
S_{B}= \int \mathrm{d}^{4}x \sqrt{-g} \left[    -\frac{1}{4}B^{\mu\nu}B_{\mu\nu} +\frac{2 \xi}{\kappa^{2}}B^{\mu}B^{\nu}R_{\mu\nu} - V(B^{\mu}B_{\mu} \mp b^{2})   \right],
\fe
where $B_{\mu\nu}=\partial_{\mu}B_{\nu}-\partial_{\nu}B_{\mu}$, $\xi$ is a positive parameter which allows the nonminimal coupling between the bumblebee field and the Ricci tensor $R_{\mu\nu}$, and $\kappa^{2}= 32\pi G$ is the gravitational coupling constant. Here, it is worth mentioning that, considering the natural units, the mass dimension of such fields and parameters are $[B^{\mu}]=1$, $[B^{\mu\nu}]=2$, $[\kappa^{2}]=-2$, $[\xi]=-2$. Next, seeking for simplicity, we adopt the smooth quadratic potential which triggers the spontaneous Lorentz symmetry breaking
\ie
V = \frac{\lambda}{2} \left(  B_{\mu}B^{\mu} \mp b^{2} \right)^{2},
\fe
where the vector $b_{\mu}$ is the vacuum expectation value of the bumblebee field $B_{\mu}$ having its minimum when $g_{\mu\nu}B^{\mu}B^{\nu} \pm b^{2}=0$. In Ref. \cite{maluf2014}, regardless torsion, the authors examined the graviton spectrum using the weak field approximation for the Einstein-Hilbert gravity in the context of Lorentz violation. For the sake of obtaining its respective Feynman propagator, we focus only on the kinetic part
\ie
\mathcal{L}_{kin}= -\frac{1}{2}h^{\mu\nu}\hat{\mathcal{O}}_{\mu\nu,\alpha\beta}h^{\alpha\beta},
\fe
where $\hat{\mathcal{O}}\indices{^\mu^\nu_\lambda_\sigma}$ is the wave operator associated to the theory. Following the definitions encountered in Ref. \cite{maluf2013}, the graviton propagator is defined as follows
\ie
\braket{0|T[h_{\mu\nu}(x)h_{\alpha\beta}(y)]|0}=D_{\mu\nu,\alpha\beta}(x-y).
\fe
Here, the main issue is finding a closed tensor algebra in order to obtain such operator $D_{\mu\nu,\alpha\beta}(x-y)$ which satisfies the Green's function 
\ie
\hat{\mathcal{O}}\indices{^\mu^\nu_\lambda_\sigma} D^{\lambda \sigma,\alpha \beta}(x-y)= i \mathcal{I}^{\mu\nu,\alpha\beta}\delta^{4}(x-y),
\fe
where $\mathcal{I}^{\mu\nu,\alpha\beta}$ plays the role of the identity operator being defined as $\mathcal{I}^{\mu\nu,\alpha\beta}=\frac{1}{2} (\eta^{\mu\alpha}\eta^{\nu\beta}+\eta^{\mu\beta}\eta^{\nu\alpha})$. Now, after many algebraic manipulations seeking the inversion of the wave operator $\hat{\mathcal{O}}\indices{^\mu^\nu_\lambda_\sigma}$, we get
\ie
\begin{split}
D_{\mu\nu,\alpha\beta}= &\frac{i}{\boxplus(k)} \left\{ \frac{N_{1}}{\kappa^{2}\xi^{2}(b \cdot k)^{2}\boxdot} \mathrm{P}^{(1)}_{\mu\nu,\alpha\beta} + \mathrm{P}^{(2)}_{\mu\nu,\alpha\beta} - \frac{1}{2} \mathrm{P}^{(0-\theta)}_{\mu\nu,\alpha\beta} + \frac{N_{4}}{2\lambda \kappa^{2}\xi^{2}(b \cdot k)^{2} \boxdot^{2}} \mathrm{P}^{(0-\omega)}_{\mu\nu,\alpha\beta} \right. \\
& \left.
+ \frac{k^{2}}{\boxdot} \Pi^{(2)}_{\mu\nu,\alpha\beta} + \frac{N_{5}}{2\xi(b \cdot k)^{2}\boxdot} \tilde{\mathrm{P}}^{(0-\theta\omega)}_{\mu\nu,\alpha\beta} + \frac{k^{2}}{\xi(b \cdot k)\boxdot}\tilde{\Pi}^{(1)}_{\mu\nu,\alpha\beta} + \frac{N_{8}}{4 \xi (b \cdot k )\boxdot} \tilde{\Pi}^{(\theta\Sigma)}_{\mu\nu,\alpha\beta}  \right.\\
& \left. - \frac{\sqrt{3}k^{2}}{2 \boxdot} \tilde{\Pi}^{(\theta\Lambda)}_{\mu\nu,\alpha\beta} + \frac{k^{4}}{2\boxdot^{2}}\tilde{\Pi}^{(\Lambda\Lambda)}_{\mu\nu,\alpha\beta} + \frac{N_{11}}{8 \xi^{2}(b \cdot k )^{2}\boxdot^{2}}\tilde{\Pi}_{\mu\nu,\alpha\beta}^{(\omega\Lambda-a)} \right.\\
& \left. +\frac{N_{12}}{2 \xi (b \cdot k)^{2}\boxdot^{2}}\tilde{\Pi}^{(\omega\Lambda-b)}_{\mu\nu,\alpha\beta} + \frac{N_{13}}{4\kappa^{2}\xi^{2}(b \cdot k)^{3}\boxdot^{2}}\tilde{\Pi}^{(\omega\Sigma)}_{\mu\nu,\alpha\beta} + \frac{N_{14}}{4\xi(b \cdot k)\boxdot^{2}}\tilde{\Pi}^{(\Lambda\Sigma)}_{\mu\nu,\alpha\beta} \right\},
\label{green}
\end{split}
\fe
with $\boxplus (k)$ and $\boxdot (k)$  being given by
\ie
\boxplus(k) = k^{2} + \xi (b \cdot k)^{2},
\label{drr1}
\fe
and
\ie
\boxdot (k) = (b\cdot k)^{2} - b^{2}k^{2},
\fe
where our attention will be devoted. Moreover, such propagator\footnote{If one is interested in any missing definitions of Eq. (\ref{green}), see Ref. \cite{maluf2014} for further details.} was verified to be physical reasonable, since it is in agreement with causality and unitarity. Nevertheless, there is no necessity of working with the full expression in our case, since all we need is fully contained in the pole of the propagator, i.e., $\boxplus (k)$. More so, it is important to mention that the pole $\boxdot (k)$ will be overlook due to the fact that it does not represent a physical mode, since it gives rise to nonunitary dispersion relation in the spacelike configuration, and it has no positive defined energy in the timelike configuration as well.

\begin{figure}[ht]
\centering
\includegraphics[width=8cm,height=5cm]{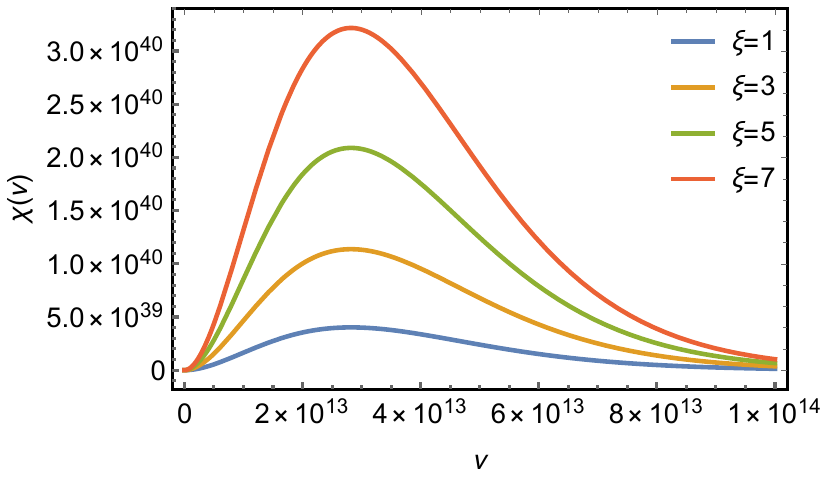}
\includegraphics[width=8cm,height=5cm]{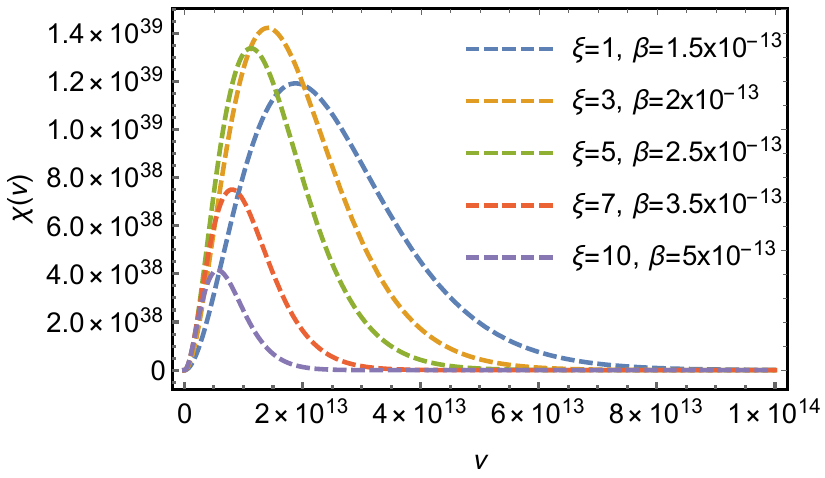}
\includegraphics[width=8cm,height=5cm]{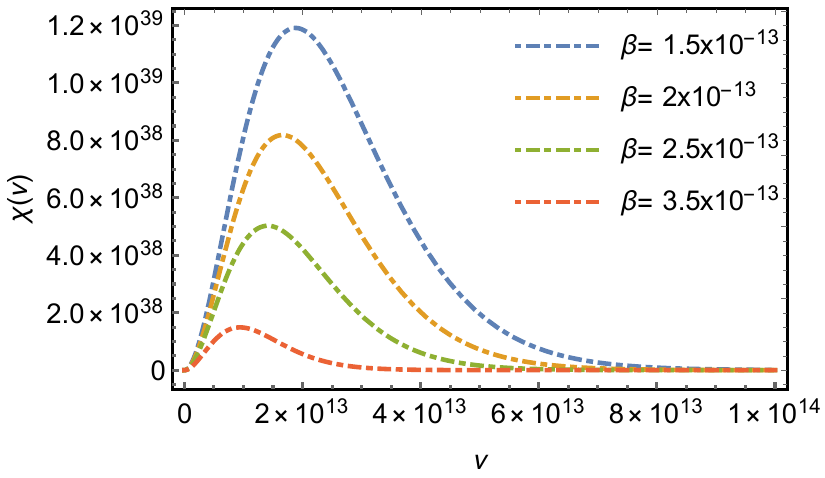}
\caption{The plots exhibit how the spectral radiance $\chi(\nu)$ changes as a function of frequency $\nu$ for different scenarios with $h=1$.}
\label{rss}
\end{figure}

\begin{figure}[ht]
\centering
\includegraphics[width=8cm,height=5cm]{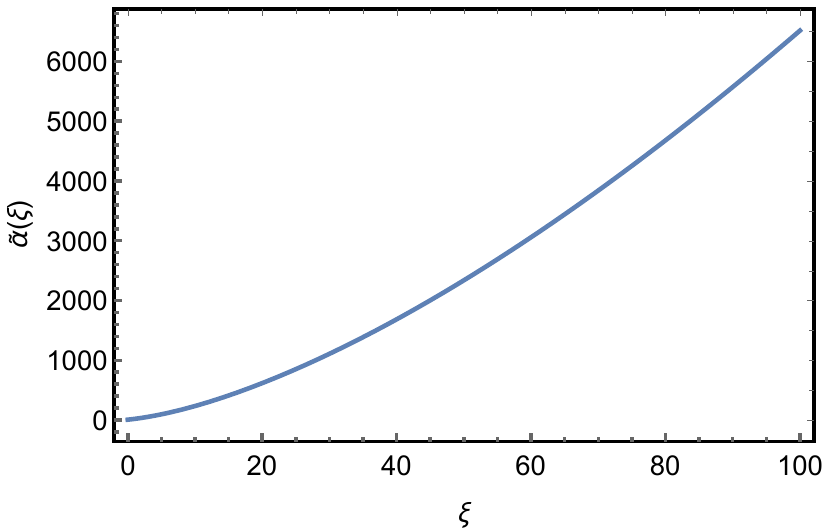}
\includegraphics[width=8cm,height=5cm]{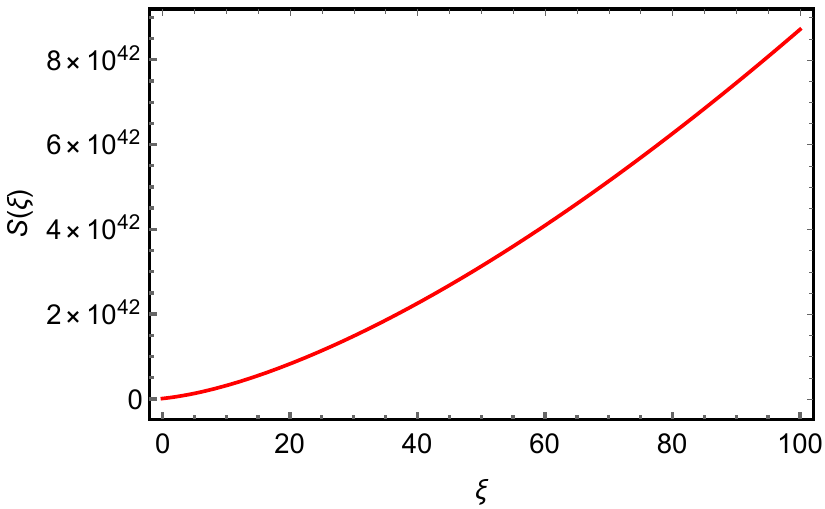}
\includegraphics[width=8cm,height=5cm]{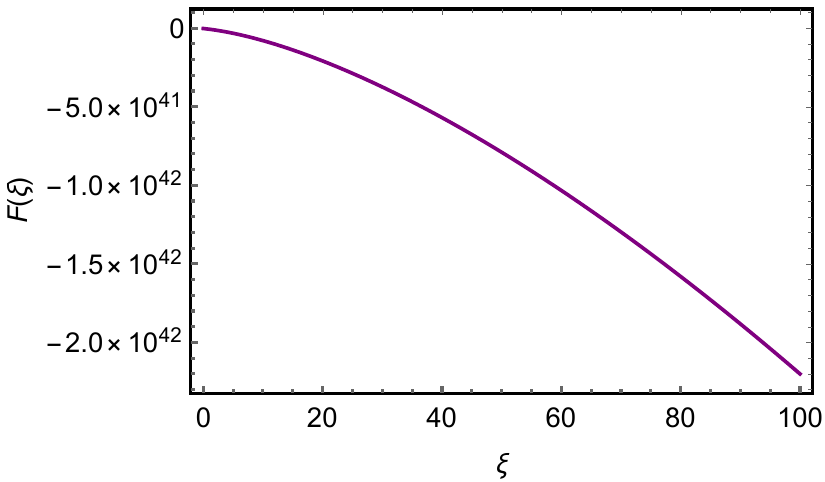}
\includegraphics[width=8cm,height=5cm]{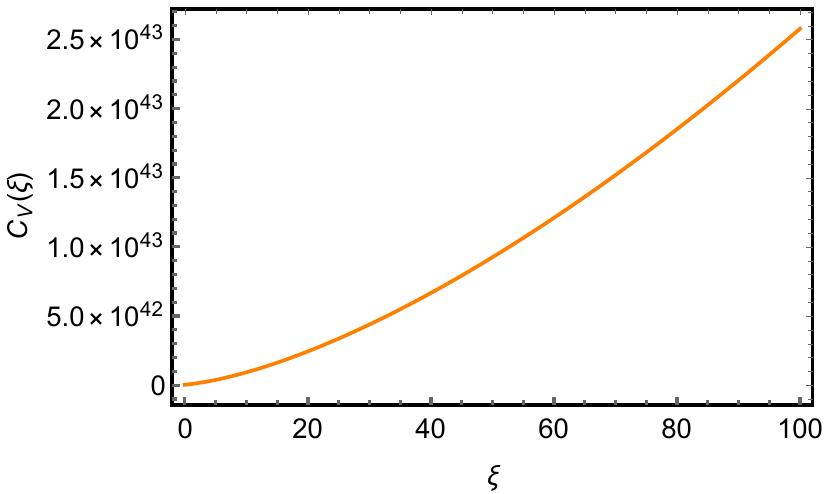}
\caption{The figure shows the correction to the so-called \textit{Stefan–Boltzmann} law represented by parameter $\tilde{\alpha}(\xi)$, the entropy $S(\xi)$, the Helmholtz free energy $F(\xi)$ and the heat capacity $C_{V}(\xi)$, considering $\kappa_{B}=1$ in the high temperature regime of the universe, namely, $\beta = 10^{-13}$ GeV.}
\label{all}
\end{figure}

In possession of this, the following calculations will be performed in order to derive all the main thermodynamic functions. For doing so, we proceed further seeking the number of the available states of the system in order to build up the so-called partition function. Here, we start with the following dispersion relation given by
\ie
k^{2}+\xi (b \cdot k)^{2} =0,
\nonumber
\fe  
and, in this sense, we take the unitary time like configuration for $b_{\mu}$, namely $b_{\mu}= (1,{\bf{0}})$, which yields the accessible states of the system written as
\ie
\Omega(\xi) = \frac{\overset{\nsim}{\Gamma}}{\pi^{2}} \int^{\infty}_{0}  (1+\xi)^{3/2}E^{2} \,\mathrm{d}E,
\fe
where $\overset{\nsim}{\Gamma}$ is the volume of the thermal reservoir. Now, let us remind here that the link between the thermal behavior and the macroscopic world is carried out by the partition function. Then, we are properly able to write it down 
\ie
\mathrm{ln}\left[ Z(\beta,\Gamma)\right] = - \frac{\overset{\nsim}{\Gamma}}{\pi^{2}} \int^{\infty}_{0}  (1+\xi)^{3/2}E^{2} \mathrm{ln} \left(  1- e^{-\beta E} \right)\mathrm{d}E,
\label{partition}
\fe
where $\beta = 1/\kappa_{B}T$ and $T$ is the temperature of the Universe. The above expression is similar to Bose-Einstein statistics but rather having a modification due to parameter $\xi$. In a straightforward manner, using the advantage of taking into account Eq.(\ref{partition}), we can obtain the thermodynamic functions per volume $\overset{\nsim}{\Gamma}$, namely, the Helmholtz free energy $F(\beta,\xi)$, the mean energy $U(\beta,\xi)$, the entropy $S(\beta,\xi)$ and the heat capacity $C_{V}(\beta,\xi)$, defined as follows:
\ie
\begin{split}
 & F(\beta,\xi)=-\frac{1}{\beta} \mathrm{ln}\left[Z(\beta,\xi)\right], \\
 & U(\beta,\xi)=-\frac{\partial}{\partial\beta} \mathrm{ln}\left[Z(\beta,\xi)\right], \\
 & S(\beta,\xi)=k_B\beta^2\frac{\partial}{\partial\beta}F(\beta,\xi), \\
 & C_V(\beta,\xi)=-k_B\beta^2\frac{\partial}{\partial\beta}U(\beta,\xi).
\label{properties}
\end{split}
\fe  
At the beginning, let us consider the mean energy
\ie
U(\beta,\xi) = \frac{1}{\pi^{2}}  \int^{\infty}_{0}  \frac{(1+\xi)^{3/2}E^{3}\,e^{-\beta E}} {\left(  1- e^{-\beta E} \right)} \mathrm{d}E, \label{meanenergy}
\fe
which follows the spectral radiance given by:
\ie
\chi(\xi,\nu) = \frac{(h\nu)^{3}(1+\xi)^{3/2}\,e^{-\beta h\nu}} {\pi^{2}\left(1- e^{-\beta h\nu} \right)},
\label{spectralradiance}
\fe
where we have regarded $E=h\nu$, as $h$ being the Planck constant and $\nu$ the frequency. The plot of the above equation is exhibited in Fig. \ref{rss} concerning three different cases when parameters $\xi$ and $\beta$ vary. This and other comments are better explained and discussed in Section \ref{Results}.
Looking toward to recover the radiation constant of the \textit{Stefan-Boltzmann} energy, i.e., $u_{S}= \alpha T^{4}$, we consider $\xi \longrightarrow 0$ leading to
\ie
\alpha = \frac{1}{\pi^{2}}  \int^{\infty}_{0}  \frac{E^{3}\,e^{-\beta E}} {\left(  1- e^{-\beta E} \right)} \mathrm{d}E = \frac{\pi^{2}}{15},
\label{radiance}
\fe
which reproduces the well-established result in the literature \cite{reif2009}. Now, for the sake of completeness, it is important to point out that hereafter, unless stated otherwise, all the following computations will be performed having the temperature $\beta= 10^{-13}$ GeV, $\kappa_{B}=1$, as well as the density per volume $\overset{\nsim}{\Gamma}$ approach. In this sense, in order to check how the coupling constant $\xi$ affects the new radiation constant and all the remaining thermodynamic functions, we proceed as follows:
\ie
\tilde{\alpha} \equiv U(\beta,\xi) \beta^{4}, \label{sbl}
\fe
and we calculate the Helmholtz free energy
\ie
F(\beta,\xi) = \frac{1}{ \pi^{2} \beta} \int^{\infty}_{0}  (1+\xi)^{3/2}E^{2}\,\mathrm{ln}\left( 1-e^{-\beta E}\right) \mathrm{d}E, \label{helmontz}
\fe
the entropy
\ie
S(\beta,\xi) = \frac{\kappa_{B}}{ \pi^{2}} \left( -\int^{\infty}_{0} (1+\xi)^{3/2}E^{2}\,\mathrm{ln}\left( 1-e^{-\beta E}\right)+ \beta\int^{\infty}_{0} \frac{(1+\xi)^{3/2}E^{3}\,e^{-\beta E}}{1-e^{-\beta E}}\right)
\mathrm{d}E, \label{entropy}
\fe
and the heat capacity
\ie
C_{V}(\beta,\xi) = \frac{\kappa_{B} \beta^{2}}{ \pi^{2}} \left(\int^{\infty}_{0} \frac{ (1+\xi)^{3/2}E^{4}\, e^{-2 \beta E}}{\left(1- e^{-\beta E}\right)^{2}}+ \int^{\infty}_{0} \frac{(1+\xi)^{3/2}E^{4} \,e^{-\beta E}}{1-e^{-\beta E}}\right) \mathrm{d}E. \label{heatcapacity}
\fe
Now, having obtained these expressions, we can solve them and their following results are displayed in Fig. \ref{all}. Notably, within the context of a linearized theory of gravity, there exists a corresponding intrinsic entropy ascribed to any distribution of gravitational radiation \cite{smolin1985} and a well-behavior conjecture having the absence of an ultraviolet catastrophe \cite{smolin1984}.

\section{Generalized model with Podolsky and Lee-Wick terms}

Recently, in the literature, the authors have proposed an effective model of higher-derivative electrodynamics in the context of Lorentz violation which studies some classical aspects regarding unitarity and causality from the propagator, i.e., it is proposed a generalized model involving anisotropic Podolsky and Lee-Wick terms \cite{casana2018}. In such reference, the advantage regarding the spin-projection operators is used \cite{scarpelli2003,adailton2}, seeking a closed algebra in order to calculate the propagator of this respective theory. For doing so, the prescription of a rank-2 symmetric tensor $D_{\beta\alpha} = (B_{\beta}C_{\alpha}-B_{\alpha}C_{\beta})/2$ (where $B_{\beta}$ and $C_{\alpha}$ are constant background four-vectors which account for Lorentz violation) has been invoked. In this sense, it was considered a more general dimension-6 higher-derivative Lagrangian was considered
\ie
\mathcal{L}= - \frac{1}{4} F^{\mu\nu}F_{\mu\nu} + \frac{\theta^{2}}{2}\partial_{\alpha}F^{\alpha\beta}\partial_{\lambda}F\indices{^\lambda_\beta} +\eta_{1}^{2}D_{\beta\alpha}\partial_{\sigma}F^{\sigma\beta}\partial_{\lambda}F^{\lambda\alpha} + \eta_{2}^{2}D^{\beta\alpha}\partial_{\sigma}F^{\sigma\lambda}\partial_{\beta}F_{\alpha\lambda} + \frac{1}{2\tilde{\xi}}(\partial_{\mu}A^{\mu})^{2} \label{La2}
\fe
where $\theta$, $\eta_{1}$ and $\eta_{2}$ are coupling constants with positive defined values and $\tilde{\xi}$ is the gauge fixing parameter to invert the wave operator associated with the Lagrangian of this theory. Besides, Eq. (\ref{La2}) leads to the corresponding propagator\footnote{Likewise in the previous section, for any missing definitions of Eq. (\ref{propagador2}), see Ref.\cite{casana2018} for further details.}
\ie
\begin{split}
\tilde{\Xi}_{\nu\alpha}(k) = - \frac{i}{k^{2} \Delta(k)} &\{ \tilde{\Gamma}(k) \Theta_{\nu\alpha} +[b'-\tilde{\xi} \Delta(k)]\Omega_{\nu\alpha} -i \tilde{F}(k) (B_{\nu}k_{\alpha}+B_{\alpha}k_{\nu}) \\
&- 2 \eta^{2}_{1}D_{\nu\alpha}k^{2}\tilde{\Pi}(k) -i \tilde{H}(k)(C_{\nu}k_{\alpha}+C_{\alpha}k_{\nu})\\
&+\eta^{4}_{1}B_{\nu}B_{\alpha}[(C \cdot k )^{2}-C^{2}k^{2}]k^{2}+ \eta^{4}_{1} C_{\nu}C_{\alpha}[(B \cdot k)^{2}-B^{2}k^{2}]k^{2}\},
\label{propagador2}
\end{split}
\fe
where $\tilde{\Gamma}(k) =\eta_{1}^{4}[(B \cdot k)^{2}-B^{2}k^{2}][(C \cdot k)^{2} -C^{2}k^{2}] - \{1 - \theta^{2}k^{2} - \eta_{1}^{2}k^{2}  (B \cdot C) +[\eta_{1}^{2}-2\eta_{2}^{2}] \\
\times(B \cdot k)(C \cdot k)  \}^{2}$ and $\Delta(k)= [ 1 - \theta^{2}k^{2} -\eta_{2}^{2} (B \cdot k)(C \cdot k)] \,\tilde{\Gamma}(k)$. In addition, Eq. (\ref{La2}) gives rise to the following dispersion relation
\ie
k^{2}\left[  1 - \theta^{2}k^{2} -\eta_{2}^{2} (B \cdot k)(C \cdot k) \right]\tilde{\Gamma}(k) =0.\label{polepropagator2}
\fe
Here, let us regard a timelike isotropic configuration characterized by $B_{\mu}= (B_{0}, {\bf{0}})$ and $C_{\mu}= (C_{0}, {\bf{0}})$. This assumption gives rise to three \textit{independent} dispersion relations to Eq. (\ref{polepropagator2}) \cite{casana2018} as follows
\ie
E_{1}^{2} = \frac{1}{1+2\eta^{2}_{2}B_{0}C_{0}/\theta^{2}}  {\bf{k}}^{2} + \frac{1}{\theta^{2}+2\eta^{2}_{2}B_{0}C_{0}}, \label{dispr2}
\fe
\ie
E_{2}^{2} = \frac{\theta^{2}}{\theta^{2}+2\eta^{2}_{2}B_{0}C_{0}}  {\bf{k}}^{2} + \frac{1}{\theta^{2}+2\eta^{2}_{2}B_{0}C_{0}}, \label{dispr3}
\fe
and
\ie
E_{3}^{2} = \frac{\theta^{2}+2\eta^{2}_{1}B_{0}C_{0}}{\theta^{2}+2\eta^{2}_{2}B_{0}C_{0}}  {\bf{k}}^{2} + \frac{1}{\theta^{2}+2\eta^{2}_{2}B_{0}C_{0}}, \label{dispr4}
\fe
where, notably, the term $1 +2\eta^{2}_{2}B_{0}C_{0}/\theta^{2}$ may be identified as a dielectric constant modifying the usual Podolsky electrodynamics. In order to perform a complete analysis of the thermal aspects of this theory displayed in Eq. (\ref{polepropagator2}), we shall examine our system considering all dispersion relations shown in Eqs. (\ref{dispr2}), (\ref{dispr3}), and (\ref{dispr4}). To do so, we consider the following approach to acquire our results, namely, $E^{2} = E_{1}^{2} + E_{2}^{2} + E_{3}^{2} $.


In possession of Eqs. (\ref{dispr2}), (\ref{dispr3}), (\ref{dispr4}) and considering a photon gas in a thermal bath, the number of available states can be derived in a straightforward way:
\ie
\bar{\Omega}(\theta,\eta_{1},\eta_{2},B_{0},C_{0}) = \frac{1}{\pi^{2}} \int_{0}^{\infty} E \left(\frac{2\theta^{2} +2\eta_{1}^{2}B_{0}C_{0}}{\theta^{2}+2\eta^{2}_{2}B_{0}C_{0} }\right)^{-3/2}  \sqrt{E^{2}- \frac{3}{\theta^{2}+2\eta^{2}_{2}B_{0}C_{0}}} \,\,\mathrm{d}E,
\fe
yielding the partition function, which may be properly written as 
\ie
\mathrm{ln}[\bar{Z}(\theta,\beta,\eta_{1},\eta_{2},B_{0},C_{0})] = - \frac{1}{\pi^{2}} \int_{0}^{\infty}E \left(\frac{2\theta^{2} + 2\eta_{1}^{2}B_{0}C_{0}}{\theta^{2}+2\eta^{2}_{2}B_{0}C_{0} }\right)^{-3/2}  \sqrt{E^{2}- \frac{3}{\theta^{2}+2\eta^{2}_{2}B_{0}C_{0}}}\, \mathrm{ln}\left( 1-e^{-\beta E} \right) \mathrm{d}E. \label{parti2}
\fe
Additionally, it is important to mention that in different contexts, many works have been made in such direction \cite{petrov2021higher,araujo2021thermal,reis2020does,adailton,pacheco2014,oliveira2019,oliveira2020}, and a Podolsky term can be generated if quantum corrections are taken into account regarding a condensation of topological defects \cite{granado2019}. Here, from Eq. (\ref{parti2}), an analogous process to calculate all those thermodynamic quantities presented in the previous section can be performed as well. In this way, we have the mean energy 
\ie
\bar{U}(\theta,\beta,\eta_{1},\eta_{2},B_{0},C_{0}) = \frac{1}{\pi^{2}} \int_{0}^{\infty}E^{2} \left(\frac{2\theta^{2} +2\eta_{1}^{2}B_{0}C_{0}}{\theta^{2}+2\eta^{2}_{2}B_{0}C_{0} }\right)^{-3/2}  \sqrt{E^{2}- \frac{3}{\theta^{2}+2\eta^{2}_{2}B_{0}C_{0}}}\left( \frac{e^{-\beta E}}{1-e^{-\beta E}}\right) \mathrm{d}E,
\label{inter2}
\fe
which follows the spectral radiance 
\ie
\bar{\chi}(\theta,\beta,\eta_{1},\eta_{2},B_{0},C_{0}) = \frac{1}{\pi^{2}} E^{2} \left(\frac{2\theta^{2} +2\eta_{1}^{2}B_{0}C_{0}}{\theta^{2}+2\eta^{2}_{2}B_{0}C_{0} }\right)^{-3/2}  \sqrt{E^{2}- \frac{3}{\theta^{2}+2\eta^{2}_{2}B_{0}C_{0}}}\left( \frac{e^{-\beta E}}{1-e^{-\beta E}}\right),
\label{sr222}
\fe
\begin{figure}[ht]
\centering
\includegraphics[width=8cm,height=4cm]{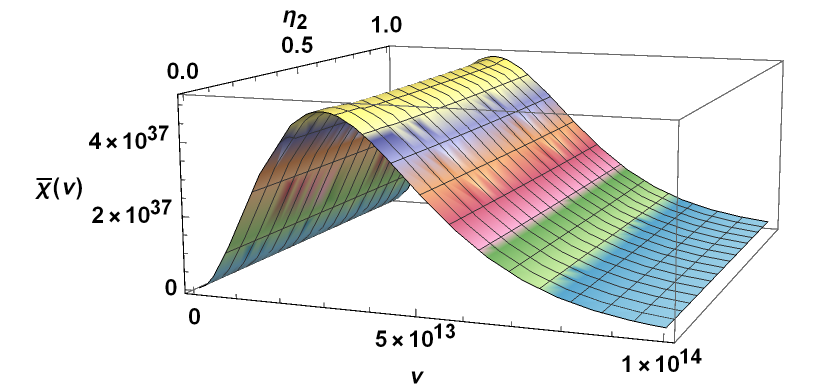}
\includegraphics[width=8cm,height=4.2cm]{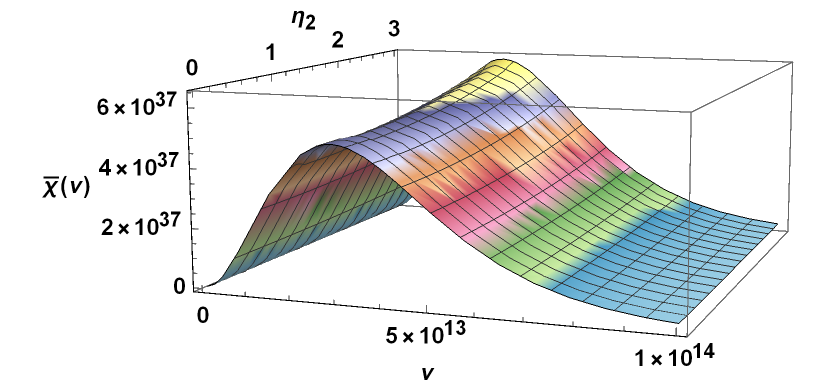}
\includegraphics[width=8cm,height=4cm]{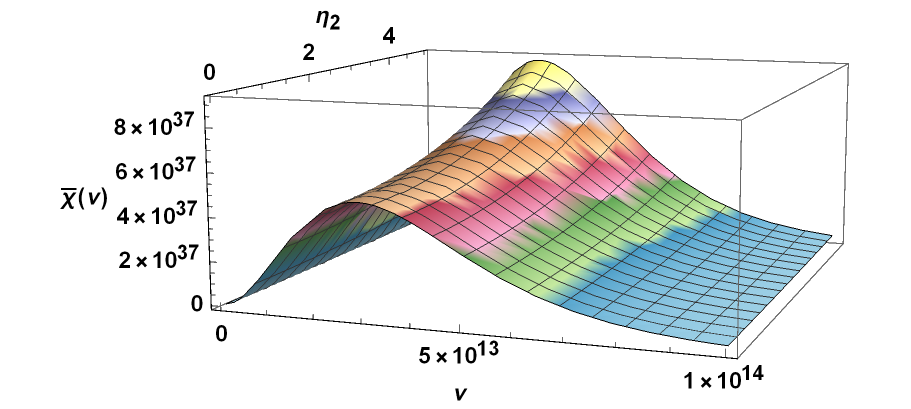}
\includegraphics[width=8cm,height=4cm]{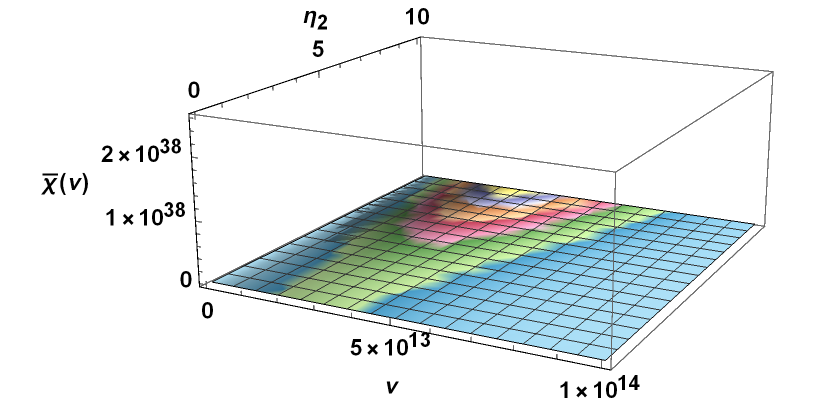}
\caption{This figure shows the behavior of the spectral radiance $\bar{\chi}(\nu)$ for different values of $\eta_{2}$ and $\nu$. We consider fixed values of $B_{0}$, $C_{0}$ and $\theta$, i.e., $\eta_{1}=B_{0}=C_{0}=1$ and $\theta=10$, in the context of the temperature in the inflationary era of the universe, i.e., $\beta = 10^{-13}$ GeV.}
\label{all12}
\end{figure}
plotted in Fig. \ref{all12} for different values of $\eta_{2}$. Again, the same procedure of inferring how the new radiation constant of the \textit{Stefan–Boltzmann} law behaves, namely $\Bar{\alpha}(\theta,\beta,\eta_{1},\eta_{2},B_{0},C_{0})$, is performed as well in what follows. Next, we derive all the remaining ones: the Helmholtz free energy
\ie
\bar{F}(\theta,\beta,\eta_{1},\eta_{2},B_{0},C_{0}) =  \frac{1}{\beta^{2}\pi^{2}} \int_{0}^{\infty}E \left(\frac{2\theta^{2} +2\eta_{1}^{2}B_{0}C_{0}}{\theta^{2}+2\eta^{2}_{2}B_{0}C_{0} }\right)^{-3/2}  \sqrt{E^{2}- \frac{3}{\theta^{2}+2\eta^{2}_{2}B_{0}C_{0}}}\, \mathrm{ln}\left( 1-e^{-\beta E} \right) \mathrm{d}E,
\label{Helmontz2}
\fe
the entropy
\ie
\begin{split}
\Bar{S}(\theta,\beta,\eta_{1},\eta_{2},B_{0},C_{0}) = &-
\frac{1}{\pi^{2}} \int^{\infty}_{0} E \left(\frac{2\theta^{2} +2\eta_{1}^{2}B_{0}C_{0}}{\theta^{2}+2\eta^{2}_{2}B_{0}C_{0} }\right)^{-3/2}  \sqrt{E^{2}- \frac{3}{\theta^{2}+2\eta^{2}_{2}B_{0}C_{0}}}\, \mathrm{ln} \left( 1 - e^{-\beta E} \right) \,\mathrm{d}E\\
& + \frac{\beta}{\pi^{2}} \int^{\infty}_{0} E^{2} \left(\frac{2\theta^{2} +2\eta_{1}^{2}B_{0}C_{0}}{\theta^{2}+2\eta^{2}_{2}B_{0}C_{0} }\right)^{-3/2}  \sqrt{E^{2}- \frac{3}{\theta^{2}+2\eta^{2}_{2}B_{0}C_{0}}} \left(\frac{e^{-\beta E}}{1-e^{-\beta E}}\right) \mathrm{d}E, \label{entropy2}
\end{split}
\fe
and, finally, the heat capacity
\ie
\begin{split}
\Bar{C}_{V}(\theta,\beta,\eta_{1},\eta_{2},B_{0},C_{0}) = &+ \frac{\beta^{2}}{\pi^{2}} \int^{\infty}_{0} E^{3} \left(\frac{2\theta^{2} +2\eta_{1}^{2}B_{0}C_{0}}{\theta^{2}+2\eta^{2}_{2}B_{0}C_{0} }\right)^{-3/2}  \sqrt{E^{2}- \frac{3}{\theta^{2}+2\eta^{2}_{2}B_{0}C_{0}}}\, \left[\frac{ e^{-2\beta E}}{\left(1-e^{-\beta E}\right)^{2}}\right]\mathrm{d}E \\
&+ \frac{\beta^{2}}{\pi^{2}} \int^{\infty}_{0} E^{3} \left(\frac{2\theta^{2} +2\eta_{1}^{2}B_{0}C_{0}}{\theta^{2}+2\eta^{2}_{2}B_{0}C_{0} }\right)^{-3/2}  \sqrt{E^{2}- \frac{3}{\theta^{2}+2\eta^{2}_{2}B_{0}C_{0}}}\, \left(\frac{e^{-\beta E}}{1-e^{-\beta E}}\right) \mathrm{d}E. \label{heatcapacity2}
\end{split}
\fe

\begin{figure}[ht]
\centering
\includegraphics[width=8cm,height=5cm]{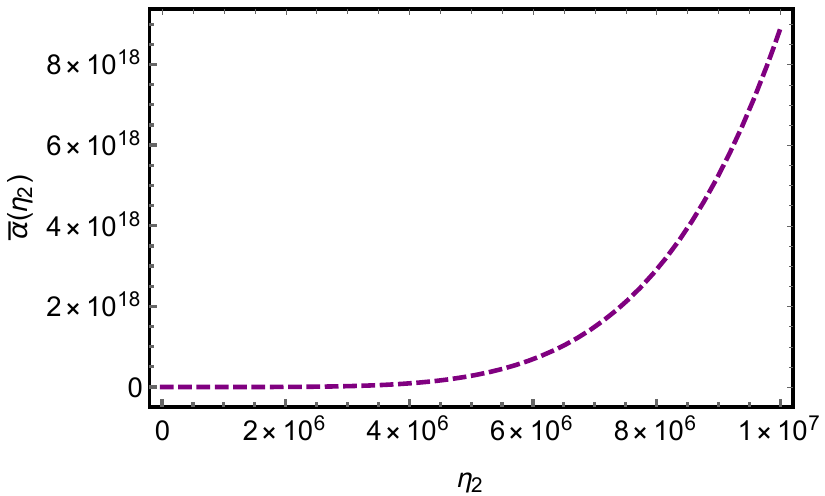}
\includegraphics[width=8cm,height=5cm]{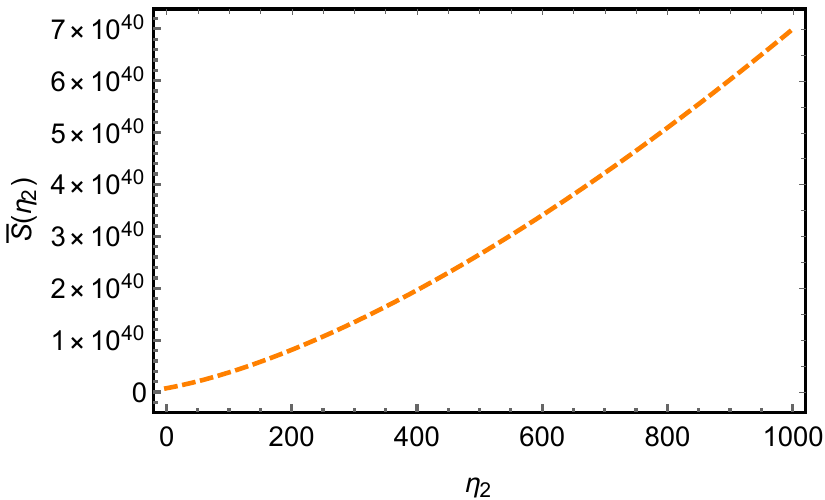}
\includegraphics[width=8cm,height=5cm]{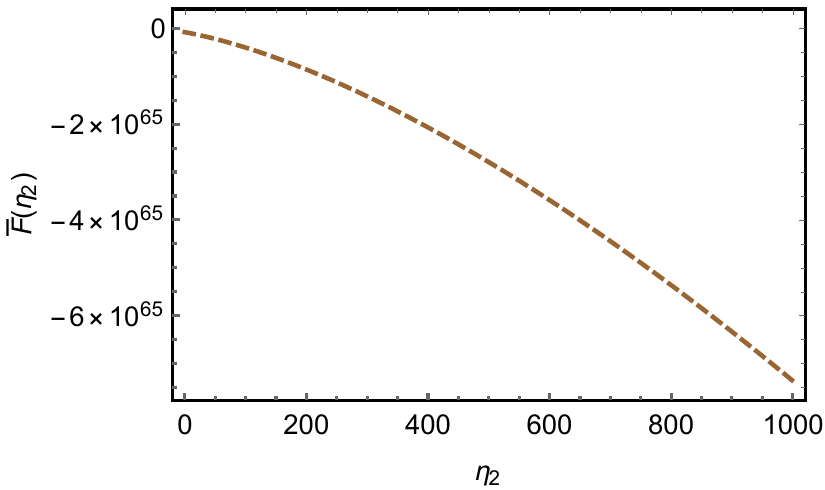}
\includegraphics[width=8cm,height=5cm]{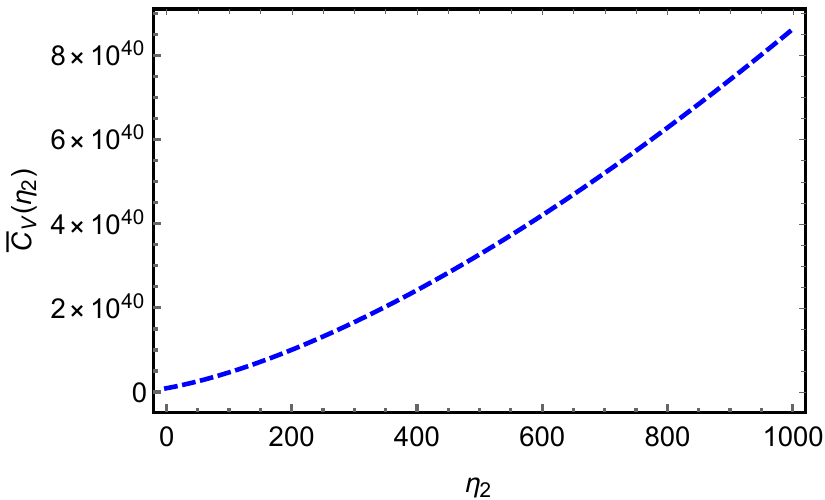}
\caption{The figure displays the correction to the so-called \textit{Stefan–Boltzmann} law represented by parameter $\Bar{\alpha}(\eta_{2})$, the entropy $\Bar{S}(\eta_{2})$, the Helmholtz free energy $\Bar{F}(\eta_{2})$ and the heat capacity $\Bar{C}_{V}(\eta_{2})$ considering $\kappa_{B}=1$ in the inflationary epoch of the universe, i.e., $\beta = 10^{-13}$ GeV.}
\label{all2}
\end{figure}

Furthermore, the graphics of these quantities are displayed in Fig. \ref{all2}. It is worth mentioning that even though there exists the appearance of a minus sign in all those square roots, as long as the positive defined values of $\theta$ and $\eta_{2}$ are considered, the theory does not possess any disturbing issues ascribed to imaginary energies. In addition, considering also the CPT-even scenario, the authors calculated the contribution to the free energy in the rotationally invariant Lorentz-violating quantum electrodynamics as well as the correction to the pressure for one-and two-loop approximations at high temperature regime \cite{gomes2010}. 

\section{Results and discussion}\label{Results}

At the beginning, we started off with the subsequent discussion regarding the thermodynamic aspects of the graviton with Lorentz violation. In this sense, we proceeded the calculations seeking the number of available states of the system which came from the given dispersion relation exhibited in Eq. (\ref{drr1}). From it, the so-called partition function was built up in Eq. (\ref{partition}) which sufficed to provide all the required thermodynamic functions, i.e., the spectral radiance $\chi(\beta,\xi)$, the mean energy $U(\beta,\xi)$, the Helmholtz free energy $F(\beta,\xi)$, the entropy $S(\beta,\xi)$ and the heat capacity $C_{V}(\beta,\xi)$.

Next, in Fig \ref{rss}, the spectral radiance was plotted for three different cases, namely, on the top left, the graphic exhibited how $\chi(\nu)$ changed as a function of $\nu$ for a fixed temperature $\beta= 10^{-13}$ GeV; on the top right, it was shown how $\chi(\nu)$ evolved for diverse values of $\xi$ and $\beta$; on the other hand, on the bottom one, the plot presented the behavior of $\chi(\nu)$ for distinct temperatures considering $\xi=1$. Besides, in Fig \ref{all}, it was displayed the modification to the \textit{Stefan–Boltzmann} law represented by parameter $\tilde{\alpha}(\xi)$ exhibiting the characteristic of a monotonically increasing function. The same behavior was also presented when one considered the entropy $S(\beta,\xi)$ and the heat capacity $C_{V}(\beta,\xi)$. However, having a different behavior from the other ones, the Helmholtz free energy $F(\beta,\xi)$ showed a monotonically decreasing function when $\xi$ started to increase.

Now, let us take into account the theory of generalized Podolsky with Lee-Wick terms. Likewise, we calculated the same thermodynamic functions for this case. In Fig. \ref{all12} is displayed how the spectral radiance evolved when $\eta_{2}$ and $\nu$ changed for fixed values of $\beta$, $B_{0}$, $C_{0}$ and $\theta$, i.e., $\beta= 10^{-13}$ GeV, $\eta_{1}=B_{0}=C_{0}=1$ and $\theta=10$ respectively. In such direction, it is worth mentioning that there was an intriguing point due to the fact that when one considered $\eta_{2} > 9$, one obtained a sudden behavior of such plot, namely, the flatness characteristic of the spectral radiance $\bar{\chi}(\eta_{2},\nu)$.

Furthermore, in Fig. \ref{all2}, the correction to the \textit{Stefan–Boltzmann} law characterized by $\Bar{\alpha}(\eta_{2})$ was shown to have an expressive positive curvature of such curve for huge values of $\eta_{2}$. This differed from the analysis accomplished by the study of the graviton modified by Lorentz violation, since the latter showed a very smooth curvature. Now, considering both the entropy $\Bar{S}(\eta_{2})$ and the heat capacity $\Bar{C}_{V}(\eta_{2})$, we verified that they presented a monotonically increasing function with a very smooth curvature when $\eta_{2}$ changed, being similar to those aspects concerning the study of the thermodynamic properties of the graviton in the context of Lorentz violation. Finally, having an analogous behavior of such theory, the Helmholtz free energy $\Bar{F}(\eta_{2})$, showed a monotonically decreasing curve when $\eta_{2}$ changed.


\section{Conclusion\label{conclusion}}

This work focused on investigating the thermodynamic properties of the graviton and a modified photon gas, which came from a generalized electrodynamics including anisotropic Podolsky and Lee-Wick terms, in a thermal reservoir when Lorentz violation is taken into account. In this direction, we determined the number of accessible states of the system which played a crucial role in obtaining the partition function. It sufficed to supply all the main thermodynamic functions, namely, spectral radiance, mean energy, Helmholtz free energy, entropy and heat capacity. Again, it is important to be noticed that the entire study was performed dealing with a high temperature regime, i.e., $\beta =10^{-13}$ GeV. Additionally, we proposed a correction to the black body radiation spectra as well as to the \textit{Stefan–Boltzmann} law in terms of $\xi$ and $\eta_{2}$.

Next, in the context of graviton thermodynamics with Lorentz violation, the parameter $\tilde{\alpha}(\xi)$ expressed the particular feature of being a monotonically increasing function. The same property was also displayed when one regarded the entropy $S(\xi)$ and the heat capacity $C_{V}(\xi)$. In contrast, being the distinct one, the Helmholtz free energy $F(\xi)$ indicated a monotonically decreasing function when parameter $\xi$ increased. 
 
On the other hand, considering the generalized Podolsky with addition of Lee-Wick terms, the $\Bar{\alpha}(\eta_{2})$ possessed significant positive curvature for huge values of $\eta_{2}$. Notably, taking into account the entropy $\Bar{S}(\eta_{2})$ and the heat capacity $\Bar{C}_{V}(\eta_{2})$, one verified that such plots had a monotonically increasing function with an attenuated curvature. In a complementary way, the Helmholtz free energy $\Bar{F}(\eta_{2})$ showed a monotonically decreasing curve. 

Lastly, the physical implications of the thermodynamic properties presented by the graviton might address some new fingerprints of a hidden physics which might be confronted with observatory data in the existence of  Lorentz violation. In such direction, these proposals can address a toy model for further studies involving gravitation and cosmology. Besides, with respect to the generalized theory of Podolsky with Lee-Wick terms, these properties might be advantageous in either forthcoming approaches regarding condensed matter physics or statistical thermodynamics. As a future perspective, examining the thermal features of very recent models, which appeared in the literature involving the Stückelberg electrodynamics modified by a Carroll-Field-Jackiw term \cite{ferreira2020}, and the graviton-dark photons in cosmological scenarios \cite{masaki2018}, seem to be interesting open questions to be investigated.


\section*{Acknowledgments}
\hspace{0.5cm}
The author would like to express his gratitude to Conselho Nacional de Desenvolvimento Cient\'{\i}fico e Tecnol\'{o}gico (CNPq) - 142412/2018-0, and CAPES-PRINT (PRINT - PROGRAMA INSTITUCIONAL DE INTERNACIONALIZAÇÃO) - 88887.508184/2020-00 the for financial support. The author also thanks L.L. Mesquita for the careful reading of this manuscript and A.Y. Petrov for the fruitful discussions and suggestions during the preparation of this work. More so, the author is also in debit with the anonymous Referee and J. A. A. S. Reis, who proposed some improvements to this manuscript pointing out crucial remarks. Last but not least, the author also acknowledges the Facultad de Física - Universitat de València and Gonzalo J. Olmo for the kind hospitality when part of this work was made.

\bibliographystyle{apsrev4-1}
\bibliography{main}

\end{document}